\documentclass{mn2e}

\newcommand{\hi}{H\,{\sc i}}

\newcommand{\etal}{et al.}

\newcommand{\Ros}{{\em ROSAT}}

\newcommand{\ltsim}{\raisebox{-1mm}{$\stackrel{<}{\sim}$}}

\newcommand{\gtsim}{\raisebox{-1mm}{$\stackrel{>}{\sim}$}}

\newcommand{\nh}{$N_{\rm H}$}

\newcommand{\eg}{e.g.}
\newcommand{\ie}{i.e.}
\newcommand{\etc}{etc.}
\def\arcm{\hbox{$^\prime$}}
\def\arcs{\arcm\hskip -0.1em\arcm}

\title[A {\em Chandra} View of the Anomalous {\em Half-Merger} NGC520]{A {\em Chandra}
View of the Anomalous {\em Half-Merger} NGC520}

\author[A.M. Read]{Andrew
M. Read$^{1}$ \\  
$^{1}$ Dept. of Physics and Astronomy, Leicester University, 
Leicester LE1 7RH, U.K. \\
(E-mail: amr30@star.le.ac.uk)} 

\date{Accepted ..............................; 
Received ..............................; 
in original form ..............................}

\begin{document}

\maketitle

\begin{abstract} 

High spatial and spectral resolution Chandra X-ray observations of the
anomalous merging galaxy NGC520, a similarly-evolved system to the
well-known Antennae galaxies, are presented here.

Of great interest is the fact that NGC520, on account of it being
supposedly due (as seen in various multi-wavelength studies) to the
result of an encounter between one gas-rich disk and one gas-poor
disk, appears in X-rays to be only 'half a merger'; Whereas a ULX lies
at the primary (SE), more-massive nucleus, no sources are seen at the
secondary nucleus. Whereas what appears to be a starburst-driven
galactic wind is seen outflowing perpendicular to the molecular disk
surrounding the primary nucleus, no such diffuse structure is seen
anywhere near the secondary nucleus. Comparing the X-ray properties
with those of other merging galaxies, including famous
gas-rich$-$gas-rich mergers such as the Mice and the Antennae, one
sees that, relative to its SFR, the number of ULXs seen within the
system is rather small. Similarly the total X-ray luminosity and the
fraction of this emission that appears diffuse are both a factor
$\sim2$ less than that expected based on NGC520's evolutionary merger
stage.

Though only half of NGC520 appears in X-rays as other mergers do,
there is still a wealth of structure and detail; 15 X-ray sources are
detected within the system, many of them showing long-term
variability, including a small number of bright ULXs that flatten the
source X-ray luminosity function to a level similar to that of
the Antennae and other mergers. Lastly, to see what appears to be a
starburst-driven diffuse galactic wind, with a spectrum entirely
consistent with that of other known galactic winds, though unusually,
emanating from only one of the nuclei, is a surprise, given that one
might have expected such structures to have distorted very quickly in
such a rapidly evolving environment. The wind is larger and more
massive than structures seen in evolutionarily earlier systems (\eg\
the Mice), but smaller and less massive than as seen in later systems
(\eg\ the Antennae), or classic starbursts. Perhaps these structures
can survive for longer than was previously thought.

\end{abstract}


\section{Introduction}

Merging and interacting are key elements in the life of galaxies, and
underpin most current theories of galaxy formation and
evolution. There are probably very few galaxies today that were not
shaped by interactions or even outright mergers. Many mergers appear
luminous in all wavebands. A very intense (L$_{\rm
bol}>10^{12}L_{\odot}$) and spatially extended burst of star formation
occurs in the evolution of most mergers. Toomre's (1977) hypothesis,
whereby elliptical galaxies might be formed from the merger of two
disk galaxies, is now generally accepted, such behaviour having been
modelled in many N-body simulations of mergers (e.g. Toomre \& Toomre
1972; Barnes 1988). During such an encounter, the conversion of
orbital to internal energy causes the two progenitor disks to sink
together and coalesce violently into a centrally condensed system. The
`Toomre sequence' (Toomre 1977) represents probably the best examples
of nearby ongoing mergers, from disk-disk systems to near-elliptical
remnants.

NGC~520 (Arp 157), the 'second brightest very disturbed galaxy in the
sky' (Arp 1987) lies seventh in the Toomre sequence, and is classified
as an intermediate-stage merger by Hibbard \& van Gorkom (1996), is as
radio and infrared bright as the famous merging system the {\em
Antennae}, and has two smaller tails as well as two nuclei and two
velocity systems in its spectra, indicative of a young merger. Some
properties of the system are given in Table~1. A distance to NGC520 of
28\,Mpc is assumed in this paper (Read \& Ponman 1998 [hereafter
RP98]; Tully 1988), hence 1\arcm\ corresponds to $\sim$8\,kpc. A brief
introduction to the system is given below, and the previous X-ray
observations are described in Sec.1.1.

\begin{table*}
\begin{center}
\begin{tabular}{|l|l|c|c|c|c|c|c|}	\hline
System &Other names & Distance & $\log{L_{B}}$ & $\log{L_{FIR}}$ &
	$L_{FIR}/L_{B}$ & $S_{60}/S_{100}$ & $\log{L_{rad}}$ \\ 
       &            & (Mpc)    &  (erg s$^{-1}$) & (erg s$^{-1}$) & 
                        &      & W\,hz$^{-1}$             \\ \hline
NGC~520    &Arp~157   & 28  & 43.87 & 44.15 & 1.900 & 0.651 & 22.24 \\
\hline
\end{tabular}
\caption[]{\small Selected properties of NGC520. Distance and optical
luminosity $L_{B}$ are taken from Tully (1988). FIR luminosity is
calculated from IRAS 60 and 100\,$\mu m$ fluxes, $S_{60}$ \&
$S_{100}$, (taken from the IRAS Point Source Catalogue), using the
expression $L_{FIR} = 3.65\times10^{5}\left[2.58 S_{60\mu m} +
S_{100\mu m}\right] D^{2} L_{\odot}$ (\eg Devereux \& Eales
1989). Radio luminosity is taken from Condon \etal\ 1990.  }
\label{table_sample}
\end{center}
\end{table*}

The nature of the peculiar system NGC~520 was once a puzzle; is it one
disturbed galaxy or two interacting galaxies? Stanford and Balcells
(1990) detected two galactic nuclei, just visible in the optical but
more clearly in the K-band. The less massive, secondary component (by
perhaps more than an order of magnitude) is the northwestern (NW)
knot, which is optically brighter than the main (primary; SE)
component. The main component is optically weak because it is seen
edge-on, and the light from its central region is absorbed by
interstellar dust in its disk (visible as a dark lane in Figure~1 of
Bernl\"{o}hr 1993). Furthermore, two hypotheses, either that the
nearby dwarf galaxy UGC~957 might be primarily responsible for the
disturbed morphology of a single galaxy in NGC~520, or that two
interacting disk systems formed NGC~520, were tested with numerical
simulations (Stanford and Balcells 1991). The simulations indicate
that NGC~520 contains two interacting disks which collided
$\sim3\times10^{8}$ years ago (UGC~957 was only involved in the
producing of the northern half of a tidal tail).

Tovmasyan and Sramek (1976) found that the compact radio source in
NGC~520 is situated in the dark lane between the two visible parts of
the system. Condon \etal\ (1982) later resolved this into a $6''$
extension, consistent with an edge-on disk, lying almost
east-west. More recent subarcsecond angular resolution observations of
the neutral gas and of the radio continuum structure at 1.4 and
1.6\,GHz (Beswick \etal\ 2003) show this $6''$ feature to be made up
of $\sim10-15$ individual clumps. Further, none of the clumps are
found to have a radio spectral index compatible with AGN, and hence
the most probable source of the radio emission is a nuclear starburst.

Millimetre-wave interferometer maps of the 2.6$\mu m$ CO emission
(Sanders \etal\ 1988) show a strong peak at the position of this radio
source; approximately $1.9\times10^{9}M_{\odot}$ of molecular gas is
concentrated in a region approximately 0.8\,kpc in size. More
recently, Yun \& Hibbard have mapped the CO $J=1\rightarrow 0$
emission, and this is seen to form an east-west ring-like structure
coincident with the radio structure. No molecular gas is seen near the
NW nucleus, or indeed elsewhere in the system, and Yun \& Hibbard
suggest that the progenitor disk surrounding the secondary nucleus was
rather gas-poor.

Much of the extranuclear regions of both galaxies within NGC~520
experienced a period of enhanced star formation $\sim3\times10^{8}$
years ago. The main sequence remnants of this burst are the A stars
whose features are evident in the optical spectra (Stanford 1991). The
putative burst within the less massive NW nucleus has returned to a
nominal level. The more massive, optically hidden SE nucleus produces
stars at a rate of $\sim0.7 M_{\odot}$\,yr$^{-1}$ and is the current
dominant source of star formation in this system. The star formation
rate within this region is $\sim35$ times higher than for an isolated
disk galaxy. This region dominates the mid-infrared flux of the
system, and probably produces most of the far-infrared flux seen in
NGC~520.

The X-ray observations of NGC520 prior to Chandra are described in
the following subsection. Section~2 describes the Chandra observations
and the data reduction techniques used. Discussion of the spatial,
spectral and temporal properties of the source and diffuse emission
components follow in Section~3, and in Section~4, the conclusions are
presented.

\subsection{Previous X-ray Observations}

NGC520 has only previously been observed in X-rays with ROSAT, and the
PSPC and HRI data were presented in RP98. The one PSPC source detected
in the vicinity of NGC~520 at $\alpha=01h24m34.76s$,
$\delta=+03d47m39.7s$, lies within $\sim5''$ of the radio source
resolved by Condon \etal\ (1982), and is coincident with the more
massive (the SE) of the two nuclei, as visible in the K-band image of
Stanford \& Balcells (1990). No source was detected at the position of
the secondary NW nucleus.

NGC~520 appeared to be a very compact X-ray source with, very
unusually for this type of system, and considering how infrared bright
it is, little in the way of diffuse emission $-$ a comparison of the
radial emission profile with the \Ros\ PSPC PSF indicated very little
emission beyond 0.6\arcm. The PSPC emission is almost consistent with
point source emission, and what diffuse emission exists, only makes up
a very small fraction of the total. Little could be said about the
spectral properties of the diffuse emission, except that it appeared 
soft.

The HRI image (RP98) showed more detail, and three sources were
detected, the most northerly centred less than 5$\arcs$ east of the
secondary (optically brighter) nucleus. This source, apparently
associated with the NW nucleus, appears to be the hardest of the
sources. A suggested extension to the east was also observed. The two
other HRI sources both appear to be soft and they follow the bright
band of optical emission down the north-eastern side of the system.

\section{Chandra observations, data reduction and results}

NGC520 was observed with Chandra on January 29th, 2003 for a total of
just over 41\,ks, with the back-illuminated ACIS-S3 CCD chip at the
focus (Observation ID: 2924). Data products, correcting for the motion
of the spacecraft and applying instrument calibrations, were produced
using the Standard Data Processing (SDP) system at the Chandra X-ray
Center (CXC). These products were then analysed using the CXC CIAO
software suite (version 3.0). 

A lightcurve extracted from a large area over the entire observation
was seen to be essentially constant and consistent with a low-level
rate, apart from a couple of high-background peaks. Screening of the
data to remove these periods (at a level of 2.44\,ct s$^{-1}$
arcmin$^{-2}$) was performed.
 
\subsection{Overall X-ray structure}

\begin{figure*}
\vspace{10cm} 
\includegraphics{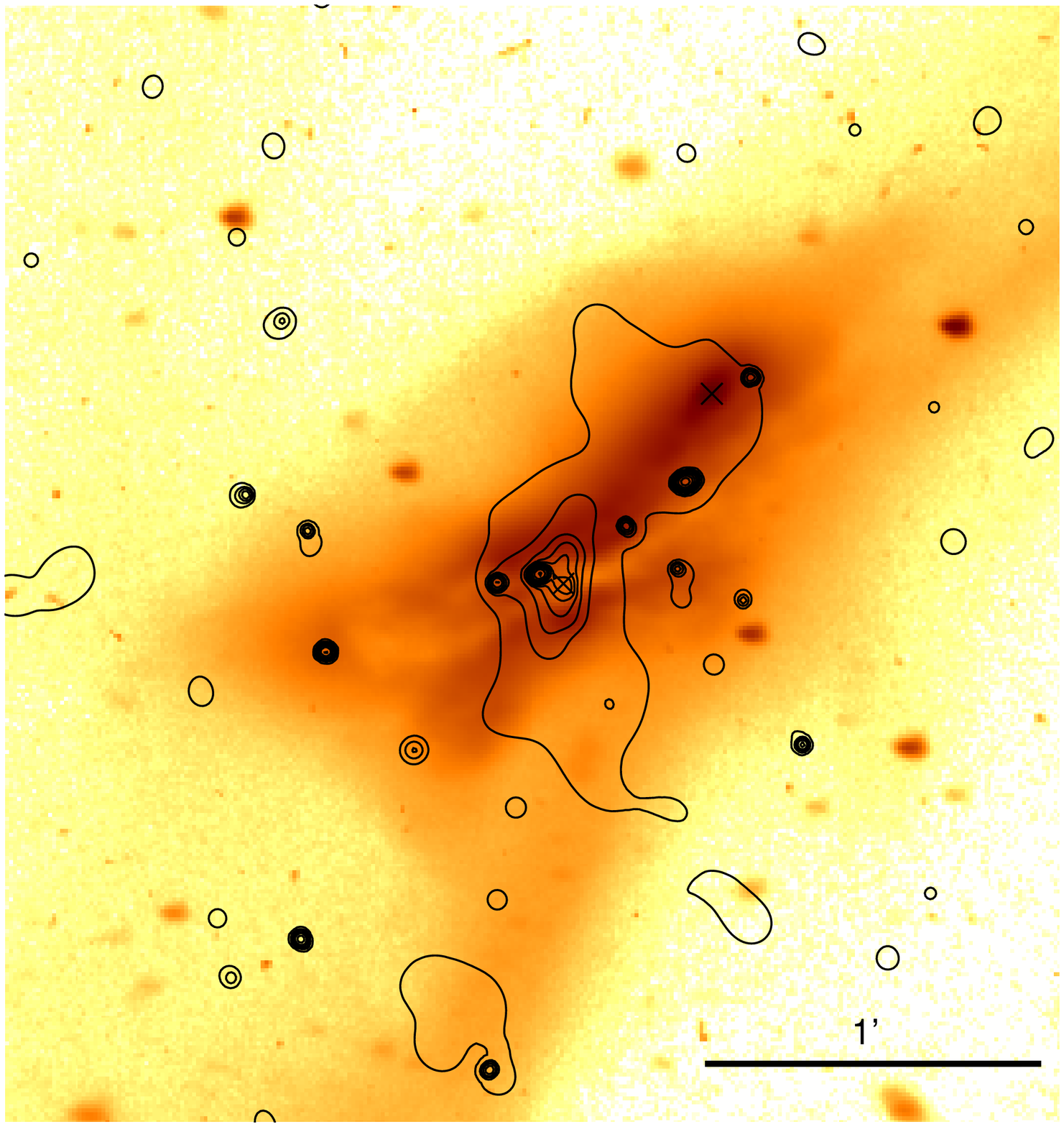} 
\includegraphics{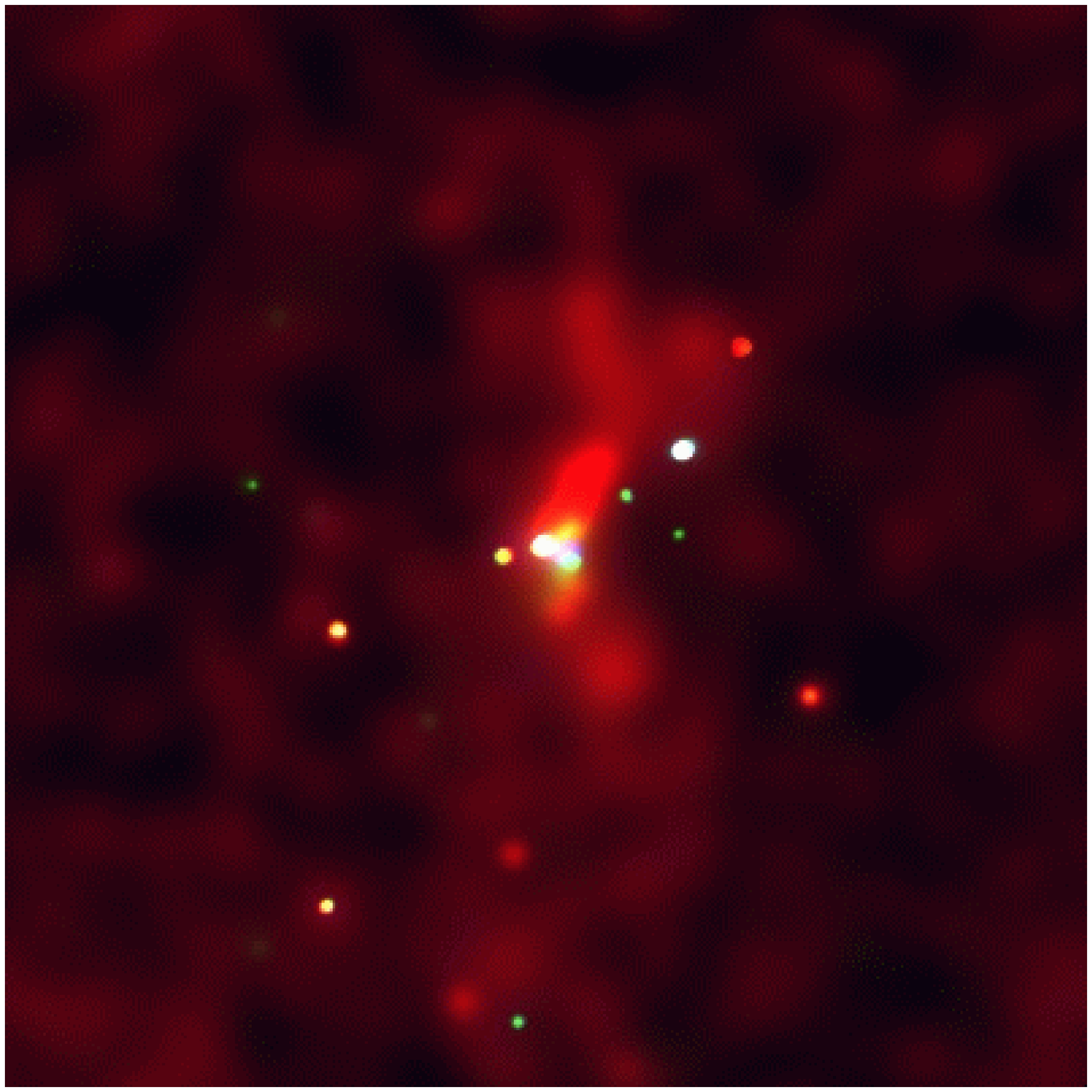} 
\caption{(Left) Contours of adaptively-smoothed (0.2$-$10\,keV)
Chandra ACIS-S X-ray emission from the field surrounding NGC520,
superimposed on the V-band KPNO 0.9\,m image from Hibbard \& van
Gorkom (1996). The X-ray contours increase by factors of two. The
crosses mark the positions of the two nuclei (from Stanford
(1991)). (Right) `True colour' X-ray image of NGC520 to the same
scale. Red corresponds to 0.2$-$0.9\,keV, green to 0.9$-$2.5\,keV and
blue to 2.5$-$10\,keV. For scale, the images are approximately 3\arcm\
to a side.}
\end{figure*}

Fig.\,1 (left) shows contours of adaptively smoothed (0.2$-$10\,keV)
Chandra ACIS-S X-ray emission from the field surrounding NGC520,
superimposed on the V-band KPNO 0.9\,m image from Hibbard \& van
Gorkom (1996). Note that the secondary (NW), less-massive nucleus is
quite visible in the optical image, whereas the primary (SE) nucleus
is hidden by the obscuring dust lane. The adaptive smoothing of the
X-ray emission attempts to adjust the smoothing kernel to obtain a
constant signal-to-noise ratio across the image.

Several things are immediately evident from the image. Many bright
point sources are visible within the optical confines of the
system. Residual, perhaps diffuse emission is seen concentrated and
centred in the dust lane at the centre of the galaxy, but extending
initially in a general north-south direction, and also following
slightly the general shape of the galaxy to the north-west.

Fig.\,1 (right) shows a true colour X-ray image, with red
corresponding to 0.2$-$0.9\,keV, green to 0.9$-$2.5\,keV and blue to
2.5$-$10\,keV. The individual adaptively-smoothed images were obtained
as described above. The image is to the same scale as Fig.\,1 (left),
and all the X-ray features are visible. A large range in spectral
hardness is seen within the point sources, and the residual, perhaps
diffuse emission is seen to be very much softer than the point source
emission.

\subsection{Point sources: spatial and spectral properties}

The CIAO tool {\em wavdetect} was used to search for point-like
sources, on scales from 1$-$16 pixels (0.5$-$8\arcs). A total of 15
sources were detected in the 0.2$-$7.5\,keV band within or close to
the optical confines of the galaxies, and their X-ray properties are
summarised in Table~2. Fig.2 shows the positions of the detected
sources.

The X-ray properties given in Table~2 are as follows; Right Ascension
and Declination (2000.0) are given in cols.\,2 and 3, together with
(col.\,4) the positional error (in arcseconds), calculated from from
the {\em wavdetect} errors in RA and Dec (no corrections on the
absolute astrometry have been applied). Net source counts (plus
errors, both from CIAO-{\em wavdetect}) are given in col.\,5, and the
source significance is given in col.\,6. Cols.\,7 and 8 give the
fitted hydrogen column and power law photon index (an 'F' indicating
that the parameter was frozen (see below for a discussion of the
spectral fitting). For those fits where a significant number of
degrees of freedom existed ($>1$; see below), this is given, together
with the best fit $\chi^{2}$ in col.\,9. Finally, (0.2$-$10\,keV)
X-ray emitted and intrinsic (\ie\ corrected for absorption)
luminosities are given in cols.\,10 and 11.

\begin{table*}
\caption[]{Sources detected by {\em wavdetect} in the 0.2$-$7.5\,keV
band within or close to the optical confines of NGC520. Columns are
described in the text. Luminosities assume a $\Gamma=1.5$ photon index
power law plus Galactic absorption, except for sources 9, 11, 12 and
15, where the spectral parameters are given. A distance of 28\,Mpc has
also been assumed.}
\begin{tabular}{lcccrrcccrr}
\noalign{\smallskip}
\hline
Src. & RA          &Dec.          &Pos.err.& Counts(err)     & Sig. & $N_{\rm H}$ & Photon & $\chi^{2}$  &\multicolumn{2}{c}{$L_{X}$ (0.2$-$10\,keV)} \\ 
     &\multicolumn{2}{c}{(2000.0)}&(arcsec)&                 &      & $10^{20}$\,cm$^{-2}$& Index  & (N$_{\rm dof}$) & \multicolumn{2}{c}{($10^{39}$\,erg s$^{-1}$)} \\ 
     &             &              &        &                 &      & & $\Gamma$&  & (emitted) & (intrinsic) \\ \hline
 7   & 01 24 37.82 & +03 46 25.9 & 0.11 &  34.7$\pm$ 6.0 & 14.2 & 3.27(F) & 1.5(F) & - & 0.75 &  0.77 \\
 8   & 01 24 37.67 & +03 47 16.1 & 0.08 &  48.2$\pm$ 7.1 & 18.5 & 3.27(F) & 1.5(F) &  - & 0.92 &  0.95 \\
 9   & 01 24 35.68 & +03 47 29.7 & 0.14 &  56.0$\pm$ 8.0 & 14.1 & 108$^{+77.8}_{-22.9}$   & 3.5$^{+0.6}_{-0.8}$  & 2.7(3) & 0.80 &  4.38 \\
10   & 01 24 35.50 & +03 46 04.9 & 0.22 &  22.4$\pm$ 5.0 &  7.9 & 3.27(F) & 1.5(F) & - & 0.74 &  0.75 \\
11   & 01 24 35.19 & +03 47 31.4 & 0.24 & 874.9$\pm$30.5 &103.2 & 40.0$^{+3.9}_{-3.5}$  & 2.4$^{+0.1}_{-0.1}$  & 74.5(73) &15.89 & 34.97 \\
12   & 01 24 34.89 & +03 47 29.5 & 0.13 & 111.6$\pm$13.2 & 12.5 & 48.4$^{+45.0}_{-21.3}$  & 1.1$^{+0.4}_{-0.2}$  & 40.4(20)& 8.89 & 10.13 \\
13   & 01 24 34.19 & +03 47 40.4 & 0.23 &  21.9$\pm$ 5.1 &  6.8 & 3.27(F) & 1.5(F) & - & 0.60 &  0.61 \\
14   & 01 24 33.53 & +03 47 33.4 & 0.32 &  11.9$\pm$ 3.7 &  4.4 & 3.27(F) & 1.5(F) & - & 0.23 &  0.25 \\
15   & 01 24 33.50 & +03 47 48.8 & 0.05 & 354.0$\pm$19.2 & 75.1 & 44.5$^{+13.7}_{-9.9}$  & 1.8$^{+0.1}_{-0.1}$  & 23.1(30) &10.00 & 13.45 \\
18   & 01 24 32.79 & +03 48 07.4 & 0.16 &  21.0$\pm$ 4.8 &  7.8 & 3.27(F) & 1.5(F) & - & 0.29 &  0.30 \\
19   & 01 24 32.73 & +03 47 29.1 & 0.35 &   8.8$\pm$ 3.2 &  3.6 & 3.27(F) & 1.5(F) & - & 0.21 &  0.22 \\
20   & 01 24 31.96 & +03 47 04.4 & 0.28 &  15.1$\pm$ 4.2 &  5.2 & 3.27(F) & 1.5(F) & - & 0.36 &  0.36 \\
23   & 01 24 40.23 & +03 45 50.0 & 0.35 &  12.0$\pm$ 3.9 &  4.1 & 3.27(F) & 1.5(F) & - & 0.17 &  0.17 \\
25   & 01 24 37.99 & +03 47 36.9 & 0.32 &   9.2$\pm$ 3.3 &  3.5 & 3.27(F) & 1.5(F) & - & 0.34 &  0.34 \\
26   & 01 24 36.60 & +03 46 59.9 & 0.31 &   8.0$\pm$ 3.0 &  3.4 & 3.27(F) & 1.5(F) & - & 0.20 &  0.22 \\
\noalign{\smallskip}
\hline
\end{tabular}
\end{table*}

Source spectra were extracted at the exact positions given by the
0.2$-$7.5\,keV detection analysis. The regions output by the detection
routines were invariably near-circles of radius $\ltsim8$ pixels
(partly due to NGC520 only occupying the very centre of the ACIS-S3
chip). Consequently a single extraction radius of 8 pixels (4\arcs)
was defined and used for all the sources in Table~1. A large area to
the SW of the system, chosen close enough to the system to minimise
effects related to the spatial variations of the CCD response, but
free of source and apparent diffuse emission, was chosen to construct
a background spectrum.

ACIS spectra were extracted using Pulse Invariant (PI) data values,
and were binned together to give a minimum of 10 counts per bin after
background subtraction. Hence $\chi^{2}$ statistics could be
used. Response matrices and ancillary response matrices were created
for each spectrum, using the latest calibration files available at the
time of writing.

Standard spectral models were fit to the spectral data using the XSPEC
spectral fitting software. Events above 7.5\,keV (of which there were
very few) and below 0.25\,keV were excluded from the fitting on the
grounds of uncertainties in the energy calibration. It is now known
that there has been a continuous degradation in the ACIS QE since
launch. A number of methods now exist within the community to correct
for this. These include the release of an XSPEC model (ACISABS) to
account for this degradation, and the existence of software (corrarf)
to correct the ancillary response files. The analysis performed in the
present paper has made use of CALDB v2.23, which does not include a
correction for the ACIS QE degradation. Hence, both the above methods
have been used here in the spectral fitting, and very similar results
were obtained. In both cases, the time since launch of the
observations (here, 1286 days) is used in the correction. Although the
calibration at energies below 1.0\,keV is believed to be uncertain,
data in this range were kept, as the statistical error on these data
points is still greater than the errors due to the uncertainties in
the calibration.

Two models, one incorporating absorption fixed at the value out of our
Galaxy (3.27$\times10^{20}$\,cm$^{-2}$) and a 5\,keV {\em mekal}
thermal plasma, the other incorporating absorption (again, fixed) and
a power-law of photon index 1.5 were fit to the source data. For the
majority of the sources, there were insufficient counts to allow the
model parameters to vary, but for sources 9, 11, 12 and 15, there were
sufficient counts ($>$50) to allow the model parameters to fit
freely. 

For 3 of these 4 cases (9, 11 and 15), F-tests showed that
statistically significant improvements in the fits were made on
freeing the parameters. Also in these 3 cases, better fits (with
reduced $\chi^{2}$\ltsim1) were obtained using a power-law model over
a thermal model, and these model parameters are quoted in Table~2. The
fit to source~12 is not too good (with a reduced $\chi^{2}$ of
$\approx$2), and this is discussed further in Section\,3.1.

The luminosities quoted in Table~2 for sources 9, 11, 12 and 15 assume
these best-fit models, while for the other sources, the model assumed
is of a fixed (Galactic) absorption plus a power-law of photon index
1.5. 

\begin{figure}
\vspace{10cm} 
\includegraphics{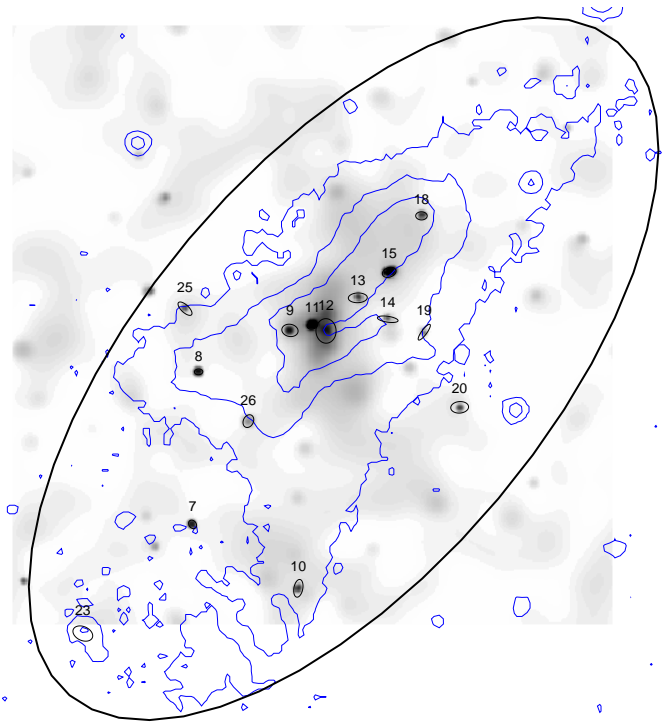} 
\caption{Adaptively-smoothed (0.2$-$10\,keV) ACIS-S image labelled
with sources detected in the 0.2$-$7.5\,keV band. The large ellipse
shows the area over which the diffuse emission spectrum was
extracted, and the blue contours show schematically the optical
confines of the system (see Fig.\,1).}
\end{figure}

\subsection{Residual emission: spatial and spectral properties}

The existence of residual, likely diffuse emission is very evident in
the figures. A spectrum from this region of apparent diffuse emission
was extracted $-$ the ellipse in Fig.\,3 shows the area over which the
spectrum was extracted, with the sources also excluded to a radius of
4\arcs. Again the spectral channels were binned together to give a
minimum of 10 counts per bin.

The spectral fitting was performed as for the point sources, using the
same models, and using both methods to correct for the degradation in
the ACIS QE. While an absorption plus power-law model was unable to
fit the data satisfactorily, a thermal model proved much better, and
the best thermal fit (using an absorption plus {\em mekal} model) is
summarised in Table~3; the absorbing column, the fitted temperature
and metallicity, the reduced $\chi^{2}$, and the emitted and intrinsic
(\ie\ total $N_{\rm H}$ absorption-corrected) X-ray luminosity is
given. The data plus best fit model is shown in Fig.\,3. A re-analysis
of the data using a grouping of 20 counts per bin gave essentially
identical results to those given here.

\begin{table*}
\caption[]{Best results of fitting a thermal model to the spectrum of
the residual emission within NGC520. Luminosities assume a distance of
28\,Mpc (see text).}
\begin{tabular}{lccccccc}
\noalign{\smallskip}
\hline

Diff. & Counts(err)    & \nh\                   & $kT$  & $Z$     & $\chi^{2}$ & \multicolumn{2}{c}{$L_{X}$ (0.2$-$10\,keV)} \\
Src.  &                & ($10^{20}$\,cm$^{-2}$) & (keV) & (solar) & (red.)     & \multicolumn{2}{c}{($10^{39}$\,erg s$^{-1}$)} \\ 
      &                &                        &       &         &            & (emitted) & (intrinsic) \\ \hline 
NGC520& 896.8$\pm$112.6& 3.27(F) & 0.58$^{+0.09}_{-0.11}$ & $<0.04$ & 0.95 & 6.40 & 8.03 \\
\noalign{\smallskip}
\hline
\end{tabular}
\end{table*}

\begin{figure}
\vspace{7cm} 
\includegraphics{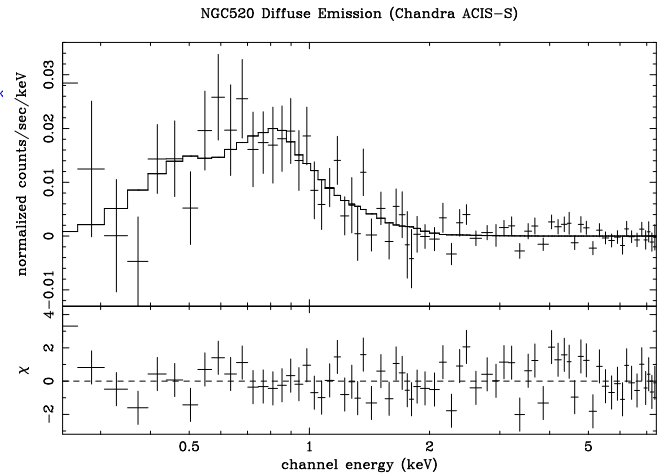} 
\caption{Data (points) plus best-fit thermal mekal model (lines) for
the residual emission within NGC520. The lower panel shows the
(data$-$model) $\Delta\chi^{2}$ values.}
\end{figure}

\section{Discussion}

\subsection{Point sources}

Within the general optical confines of NGC520 (\ie\ within the ellipse
shown in Fig.\,2), 15 sources are detected with the Chandra ACIS-S
instrument, down to a (0.2$-$10.0\,keV) detection limit of
$\approx1.7\times10^{38}$\,erg s$^{-1}$. Previous detections of X-ray
sources include the one source (P1) detected with the ROSAT PSPC, and
the three sources (H1$-$H3) detected by the ROSAT HRI (RP98). One can
use the the logN$-$logS relations of Rosati \etal\ (2002) to estimate
the expected number of background sources not physically associated
with NGC520. At most, perhaps 2 background sources are expected over
the ellipse covered by Fig.\,2 at the detection limit seen here.

Source 12 is bright (though only the third brightest source),
spectrally quite hard (as indicated by the blue-white appearance in
the RGB plot - Fig.\,1 [right]), and is coincident ($<0.5$\arcs) with
the primary nucleus, \ie\ the Condon \etal\ (1982) radio source and
the Yun \& Hibbard (2001) CO feature. It lies at the centre of the
brightest part of the much softer diffuse X-ray emission (discussed in
the next subsection). The central region of NGC520 is shown more
clearly in Fig.\,4, where the adaptively-smoothed (0.2$-$10\,keV)
Chandra ACIS-S X-ray emission (from Fig.\,1 [left]), the
velocity-integrated CO(1$-$0) map of Yun \& Hibbard (2001), the
H$\alpha$ emission from Yun \& Hibbard (2001), the Stanford (1991)
primary nucleus position, the Condon \etal\ (1982) 6\arcs\ radio
source and the positions of the main 1.4\,GHz components resolved
within the 6\arcs\ Condon feature by Beswick \etal\ (2003) are all
shown together. In actuality, Chandra source 12 lies closest to
($<0.2$\arcs) the brightest and central-most of the Beswick \etal\
(2003) radio features $-$ their feature 6. Taking also into account
uncertainties in Chandra's absolute astrometry ($\approx 0.7$\arcs for
on-axis, isolated point sources), then the Chandra source 12 error
circle could also just encompass radio feature~7 (the 2nd brightest
feature of Beswick \etal\ 2003). Chandra source 12 is undoubtedly
associated with the primary nucleus of NGC520. A spectral fit to
source 12's spectrum, but using a model incorporating an extra
component to represent the soft, thermal diffuse emission (with fixed
parameters as given in Table~3), results in a far better fit to the
source 12 data (with a reduction in reduced $\chi^{2}$ of
$\gtsim$0.5). An F-test shows this improvement to be statistically
significant at over 97\% confidence. Here, the hard component (assumed
due to the actual source 12 itself) is better represented by a
power-law model ($N_{H}$=$2.3\times10^{22}$\,cm$^{-2}$, $\Gamma$=2.15)
than a thermal model ($N_{H}$=$2.5\times10^{22}$\,cm$^{-2}$,
$kT$=3.1\,keV). In both cases, the hard (\ie\ the point source)
component makes up $\approx90$\% of the total emitted (0.2$-$10\,keV)
flux from source 12.

Source 11 is the brightest of the X-ray sources and lies at the
easternmost edge of the radio/CO disk. Though it could be enclosed
within the {\em ROSAT} H3 and H2/P1 sources and the complex and
confused PSPC/HRI emission surrounding the primary nucleus, this
appears not too likely. Source 15, the second brightest source, does
not have a direct counterpart in the HRI (or the relatively poor
positional resolution PSPC) observations, and this would indicate that
source 15 (and likely source 11) are transient in nature. Both sources
are well fit with single component power-law models (Table~2).

Sources 8 and 9 are both fairly bright and fairly
soft/medium-temperature sources. Source 9 is associated with H3, and
source 8 may have a low-significance HRI counterpart. Of the
remaining, lower-luminosity sources, source 18 lies closest to the
secondary (NW) nucleus (visible in the V-band image of Fig.\,1
[left]), but still lies some 7\arcs\ distant $-$ hence, interestingly,
no X-ray emission is coincident with the secondary nucleus, and the
HRI source (H1) tentatively associated (RP98) with the NW nucleus
appears to have vanished. Given the astrometric uncertainties in the
{\em ROSAT} position it is unlikely that source 8 is coincident with
H1.

Beswick \etal\ (2003) believe that two different components of gas
(neutral gas and ionised gas), with very different velocity
characteristics, can be sampled within NGC520, and these components
are located at different distances from the primary nucleus. Whereas
the velocity structure of the neutral ISM components within the
central regions of the primary nucleus (Fig.\,4) are probably more
characteristic of the rotational motion of the progenitor galaxy
nucleus, the very different velocities observed (\eg\ Stockton \&
Bertola 1980; Bernl\"{o}hr 1993) for the ionised gas located a few kpc
from the primary nucleus have likely been imparted on the gas by the
merger event. It is very interesting to note that that there are many
bright sources (notably 9, 11 \& 13) that lie in this zone where the
two velocity systems are colliding, a region where strong shocks and
large compressions of gas are expected to occur. Indeed, the H$\alpha$
emission (Yun \& Hibbard 2001) indicates that a good deal of
star-formation is certainly taking place on the eastern side of the
zone where the two velocity systems are colliding, in the region
occupied by sources 9 and 11.

\begin{figure}
\vspace{9cm} 
\includegraphics{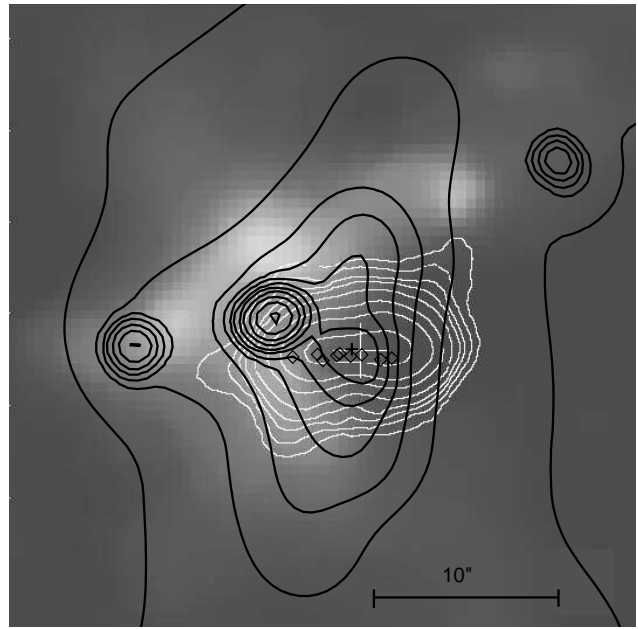} 
\caption{The central region of NGC520: Black contours show the
contours of adaptively-smoothed (0.2$-$10\,keV) Chandra ACIS-S X-ray
emission (from Fig.\,1 [left]). White contours show the
velocity-integrated CO(1$-$0) map of Yun \& Hibbard (2001). The small
black cross shows the position of the primary nucleus from Stanford
(1991). The large white cross shows the centre of the 6\arcs\ radio
source of Condon \etal\ (1982), and the diamonds show the positions of
the nine main 1.4\,GHz components resolved within the 6\arcs\ Condon
feature by Beswick \etal\ (2003). The underlying grey-scale shows the
H$\alpha$ image emission from Yun \& Hibbard (2001). The image is
$\approx33$\arcs\ across.}
\end{figure}

The cumulative X-ray luminosity function (XLF) of the sources detected
within NGC520 is shown in Fig.\,5. Plotted is the (log of the) number
of sources above a given X-ray luminosity versus (log of the) X-ray
luminosity. The functions for the intrinsic and emitted values (see
Table 2) are given. While simple regression fitting of a linear
function to $\log{N}$ against $\log{L_{X}}$ leads to a slope of 0.45
(usage of the emitted $L_{X}$ values leads to a slope of 0.50), a more
appropriate approach, and one where the errors of the Poissonian
statistics of the data are better reflected, is to fit the
differential luminosity function, using the method of Crawford \etal\
(1970). This gives more realistic errors on the fitted slope. Usage of
this method gives slopes of 0.58$\pm$0.15 for the intrinsic $L_{X}$
values (and 0.66$\pm$0.17 for the emitted $L_{X}$ values).

Colbert \etal\ (2003) have found that, while ellipticals have very
steep XLF slopes (-1.41$\pm$0.38), the slopes for spirals are much
flatter (-0.79$\pm$0.24), and for merging galaxies and irregular
galaxies, the slope is even flatter still (-0.65$\pm$0.16). NGC520
therefore has a very flat XLF, consistent with it being a highly
irregular, merging galaxy. NGC520 in fact has an XLF slope similar to
that of the {\em Antennae} (-0.47$\pm$0.05; Zezas \& Fabbiano 2002), a
system with a very similar far-infrared luminosity, and a system at a
similar to slightly later evolutionary stage.

The number of Ultraluminous X-ray Sources (ULXs), \ie\ sources with
$L_{X}$\gtsim$10^{39}$\,erg s$^{-1}$, in NGC520 is around 3$-$4 (\ie\
sources 11, 12 and 15, with possibly source 9). Were it not for these
sources, the XLF would appear rather steeper than it does in Fig.5,
looking more like that of normal spirals. The 3 definite ULXs in fact,
make up most of total X-ray flux, accounting for 85\% of the point
source emission, and 73\% of the total emission.

There is evidence to suggest that galaxies with a greater star
formation rate contain a greater number of ULX's (Swartz \etal\ 2004),
and it is these sources that flatten the XLF. Brassington \etal\
(2004) have presented a relationship between the number of ULXs and
the far-infrared luminosity for a small number of interacting and
merging galaxies: $\log{N({\rm ULX})} \propto \log{L_{\rm
FIR}^{0.18\pm0.13}}$. NGC520 appears unusual in that it does not fit
into this picture at all, having a large $L_{\rm FIR}$ value, very
like that of \eg\ the {\em Mice} or the {\em Antennae}, but a
relatively small number of ULXs (3$-$4). One would expect, based on
the number of ULXs per unit far-infrared luminosity, there to be about
10 ULXs in NGC520, if it were similar to the {\em Antennae}. 

Furthermore, Brassington \etal\ (2004) observe a relationship between
the integrated luminosity of the ULX sources and the far-infrared
luminosity in their merging galaxy sample: $\log{L_{\rm ULX}} \propto
\log{L_{\rm FIR}^{0.54\pm0.04}}$.  The small number of NGC520 ULXs are
all bright, but even so, NGC520 lies significantly below the observed
$L_{\rm ULX}-L_{\rm FIR}$ relationship.

\begin{figure}
\vspace{10cm} 
\includegraphics{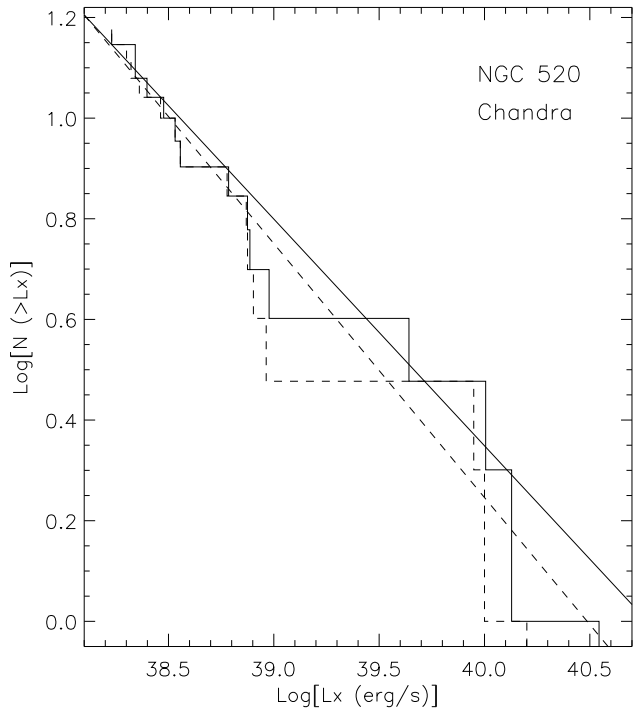} 
\caption{The cumulative luminosity function of the discrete sources
detected within NGC520. Plotted is the log of the number of sources
above a given X-ray luminosity versus (log of the) X-ray
luminosity. The functions for the intrinsic (solid line) and emitted
(dashed line) values (see Table 2) are given. The straight solid and
dashed lines show the best regression fits to the two functions (see text).}
\end{figure}

\subsection{Diffuse emission}

The residual emission, given its structure and its spectrally soft,
single-component nature, is very likely predominantly due to genuinely
diffuse hot (0.58\,keV) gas (though there will be some contribution
from unresolved sources).

It appears centred on Chandra source 12, at the precise site of the
primary, more massive, optically-obscured nucleus, a nucleus believed
not to harbour any AGN, but believed to be a starburst-powered nucleus
(Beswick \etal\ 2003). The diffuse emission is seen to extend in a
bipolar form initially to the north and to the south, i.e. directly
perpendicular to the molecular disk (Yun \& Hibbard 2001) surrounding
the primary nucleus, a disk also observed in the radio (Condon \etal\
1982; Beswick \etal\ 2003). This north-south extension is seen to be
quite bright for quite a distance ($\sim 2$\,kpc from the nucleus),
and then becomes somewhat weaker, perhaps more so to the south,
reaching a final extent of perhaps $\sim 7$\,kpc (to the north) and
$\sim 5.5$\,kpc (to the south). It is quite unlikely that the
north-south extension could be predominantly due to absorption by cold
disk gas, as the CO gas exists in only a small ($\sim12$\arcs)
centralised region, and the rather diffuse neutral H{\rm I} gas (Yun
\& Hibbard 2001) pervades much of the system, and extends in a more
SE-NW direction, in a direction roughly connecting the two nuclei. It
is very interesting to note that there appears to be no significant
enhancement in the diffuse emission, nor in the X-ray emission in
general, in the vicinity of the secondary nucleus.

All these points, both spatial and spectral, are very suggestive of
the diffuse structure being due predominantly to a small version of a
starburst-driven wind from solely the primary (SE) nucleus. Further
evidence is suggested by the ionised H$\alpha$ emission (visible in
Fig.\,4), following roughly the inner structure of the diffuse X-ray
feature. Yun \& Hibbard (2001) conclude that this H$\alpha$ emission
is likely dominated by the 'starburst-driven ionised wind' escaping to
the north and south. Interestingly, there is no evidence whatsoever
for any similar structure in the secondary (NW) nucleus. Classic
winds, such as those seen in famous nearby starburst galaxies such as
M82 and NGC253 (\eg\ Stevens, Read \& Bravo-Guerrero 2002; Strickland
\etal\ 2002) are rather isolated, and to see what appears to be a
bipolar wind in even one member of a strong, rapidly-evolving
interacting pair such as here in NGC520, is rather surprising. It is
believed (Read 2003) that the very beginning of starburst-driven hot
gaseous outflows in full-blown disk-disk mergers has been seen in the
{\em Mice}. One might have thought that later systems would have
evolved to such a degree, that any classic starburst wind or winds
(were they to have existed), would have been distorted out of
recognition, and indeed, Chandra observations of the Antennae (a
system believed to be post-Mice, but more like at the evolutionary
stage of NGC520) show a great deal of hot diffuse gas, but it has
become all-pervasive, extending beyond the stellar bodies of the
galaxies (Fabbiano \etal\ 2002).

One can infer mean physical properties of the hot gas around the
northern and southern galaxies once some assumptions have been made
regarding the geometry of the diffuse emission. Two models have been
used here. A conservative model assumes the gas here to be contained
in an elliptical bubble (the ellipse as shown in Fig.\,2 and the third
[line-of-sight] axis assumed equal ['radius' 9.2\,kpc] to the short
axis of the ellipse). A more stringent, and probably more realistic
model roughly follows the outer contours of the X-ray emission and
assumes the gas to be contained in a spherical bubble of radius
45\arcs\ (corresponding to 6.1\,kpc). Using these volumes, the fitted
emission measure $\eta n^{2}_{e} V$ (where $\eta$ is the `filling
factor' - the fraction of the total volume $V$ which is occupied by
the emitting gas) can be used to infer the mean electron density
$n_{e}$, and hence, assuming a plasma composition of hydrogen ions and
electrons, the total mass $M_{\mbox{\small gas}}$ and thermal energy
of the gas $E_{\mbox{\small th}}$. Approximate values of the cooling
time $t_{\mbox{\small cool}}$ of the hot gas, and also the mass
cooling rate $\dot{M}_{\mbox{\small cool}}$ can also be
calculated. The resulting gas parameters for both models are listed in
Table\,4.

\begin{table*}
\begin{center}
\begin{tabular}{cccccccc}    \hline
Diff. & $kT$ & r & $n_{e}$ & $M_{\mbox{\small gas}}$ &
$E_{\mbox{\small th}}$ & $t_{\mbox{\small cool}}$ &
$\dot{M}_{\mbox{\small cool}}$ \\
Model &(keV) &(kpc)&(cm$^{-3}$)& ($M_{\odot}$)           & (erg)                  & 
(Myr)                    & ($M_{\odot}$ yr$^{-1}$)        \\
      &      &     &($\times1/\sqrt{\eta}$) & ($\times\sqrt{\eta}$) &  ($\times\sqrt{\eta}$)   & 
($\times\sqrt{\eta}$)    & \\ \hline

Large Ellipse&0.58&9.2 & 6.9$\times10^{-3}$& 7.1$\times10^{8}$ & 1.9$\times10^{57}$ & 7590 & 0.094 \\ 

Small Bubble &0.58&6.1 & 1.9$\times10^{-2}$& 2.8$\times10^{8}$ & 7.6$\times10^{56}$ & 3560 & 0.079 \\

\hline
\end{tabular}
\caption{Values of physical parameters for the diffuse gas associated
with NGC520. The values quoted are for a large ellipsoidal bubble
model and a small spherical bubble model (see text).$\eta$ is the
filling factor of the gas.}
\label{table_gas}
\end{center}
\end{table*}

Comparing the diffuse gas parameters in Table~4 with those for
isolated normal and starburst galaxies (Read, Ponman \& Strickland
1997), and for merging galaxies (RP98; Fabbiano \etal\ 2002; Read
2003), one can say that the single diffuse outflow observed in NGC520
is larger than the outflows seen in the earlier-stage {\em Mice}
($2-3$\,kpc), but is not as large as the the classic winds of M82 and
NGC253 (9$-$14\,kpc), nor quite as large as the outflowing,
non-collimated turbulent ISM of the Antennae ($\approx$8\,kpc). There
appears to be more diffuse gas within the NGC520 system than in the
{\em Mice}, but this gas mass, while possibly comparable to that of
NGC253 and of order a half that seen in M82, is only 10-20\% of that
seen in the {\em Antennae}. 

The injection rate of mechanical energy, or power injected into a
particular galaxy by supernovae, can be estimated by the method
employed in Read \& Ponman (1995), and this value for NGC520 is seen
to be very similar to that of the {\em Antennae} and M82. The amount
of hot gas in NGC520, expressed as a fraction of input energy
therefore, is rather small. Note that, as the diffuse gas has a very
long cooling time, no significant fraction of the gas has had time to
cool.

Though it is believed that the hot ISM of starburst galaxies has a
multi-temperature structure, the single temperature obtained from the
spectral fitting of the diffuse spectrum ($\approx$0.58\,keV) is
entirely consistent with the range obtained for other starburst and
merging galaxies.

\subsection{X-ray emission from NGC520}

NGC520 is very far-infrared bright, and consequently has a large value
of $L_{\rm FIR}/L_{\rm B}$. It also has a large far-infrared
temperature, $S_{60}/S_{100}$, and both these are indicative of NGC520
having a high SFR and being very active, more active than the famous
interacting systems The {\em Mice} and The {\em Antennae}.

The total emitted 0.2$-$10\,keV X-ray luminosity of NGC520, assuming a
distance of 28\,Mpc is 4.74$\times10^{40}$\,erg s$^{-1}$, of which the
diffuse emission makes up 13.5\% (the equivalent intrinsic
absorption-corrected values are 7.64$\times10^{40}$\,erg s$^{-1}$ and
$f_{\rm diff}=10.5$\%).

The value of $L_{X}$ quoted here may initially appear at odds with the
low value ($\approx$1$\times10^{40}$\,erg s${-1}$) given in RP98. There
a few points here to note: The energy bands over which the
luminosities are calculated are different. The RP98 value is based on
the results of spectrally fitting the PSPC data, and this analysis was
prone to difficulties and the spectral parameters obtained had large
errors. In actuality, using PIMMS to predict the Chandra ACIS-S count
rate, based purely on the HRI count-rate, allowing for the
different energy ranges, and using a much more appropriate spectral
model (a $\Gamma=$1.5 power-law) does give a similar result to what we
actually observe with the ACIS-S (to within 2\%). Lastly, there is
quite evidently a lot of variability going on within the point
sources, so we should expect some variation in $L_{X}$ between the
observations.

As such, the larger X-ray luminosity inferred here for NGC520,
indicates that it is not as anomalous as was first thought in RP98 $-$
both its position in the $L_{X}$$-$$L_{FIR}$ plane (RP98 Fig.\,14) and
its position in the $L_{X}/L_{B}$-versus-age plane (RP98 Fig.\,13)
appear closer in line with the other similar-stage mergers than as
depicted in RP98. However, while taking into account the points raised
in the previous paragraph, NGC520 still appears somewhat underluminous
in X-rays, for its merger stage, perhaps by a factor of up to 2.

Yun \& Hibbard (2001) discovered that, though a dense concentration of
molecular gas is seen at the primary (SE) nucleus, none is seen at the
secondary (NW) nucleus, nor along the region bridging the two
nuclei. Hibbard \& van Gorkom's (1996) analysis of the stellar and
\hi\ features in NGC520 indicated that NGC520 is the product of an
encounter between a gas-rich disk and a gas-poor disk (\eg\ an S0 and
Sa galaxy). The absence here of molecular gas in the NW nucleus can be
explained if little molecular gas had been compressed towards the
nucleus during the merger process, there being little gas in the
progenitor disk to start with. Similar analyses have concluded that
systems such as the {\em Mice} and the {\em Antennae} are each the
product of encounters between two gas-rich systems.

This initial lack of gas in the NW progenitor is a very likely
explanation of several facets of the X-ray emission discussed
here. That there was no gas in the disk meant that no gas could be
funnelled towards the NW nucleus during the interaction. Consequently,
little in the way of a starburst could take place at the NW nucleus,
and no collections of supernovae could collect and expand outwards in
the form of a starburst-driven wind or of fountains or chimneys
\etc. $-$ there would be, as is observed, no diffuse X-ray emission
surrounding the NW nucleus. The percentage of the total X-ray emission
in NGC520 observed to be diffuse ($f_{\rm diff}$) is low
($\sim13.5$\%), when compared with roughly similar stage systems (\eg\
the Antennae [45\%]; Fabbiano \etal\ 2002), and even when compared
with very early-stage interacting systems (\eg\ Arp270 [29\%];
Brassington \etal\ 2004). Were the progenitor disk that formed the NW
nucleus initially gas-rich, then $f_{\rm diff}$ for NGC520 would
certainly be larger (though not by a factor of two), and perhaps more
in line with other merging systems.

Further, as discussed earlier, the number of ULXs observed is very low
when compared to other mergers. It is well known (\eg\ Roberts \etal\
2002) that ULXs are more prevalent in regions of enhanced
star-formation. If a starburst were to have occurred in the NW nucleus
similar to that in the SE nucleus, then the number of ULXs within the
NGC520 system may well have been more like that of similar merger
systems.

Lastly, the arguments raised here could increase the total X-ray
luminosity, perhaps by the factor of $\sim$2 required, bringing NGC520
more in-line with its expected position in the X-ray evolution of
merging galaxies.

\section{Conclusions}

Presented here are high spatial and spectral resolution Chandra ACIS-S
X-ray observations of the interacting galaxy system, NGC520, a system
at a similar or slightly later evolutionary stage as the {\em Antennae},
\ie\ more evolved than the {\em Mice} but not as evolved as the
ultraluminous contact mergers, such as Arp220. The primary results can
be summarised as follows:

\begin{itemize}

\item What appears to possibly be a starburst-driven galactic wind has
been discovered, outflowing perpendicular to the molecular disk
surrounding the primary, optically-obscured SE nucleus. No such
structure, nor any significant diffuse feature of any kind is visible
elsewhere in the system, and is notably absent from around the
secondary NW nucleus. Fitting of this diffuse emission yields a
temperature entirely consistent with well-known nearby galactic winds.
In terms of extent and mass, the wind appears larger and more massive
than the winds of earlier-stage mergers, but is only at most of order
the size and mass of the winds seen in the smaller nearby starbursts,
and is rather much smaller and less massive than the diffuse structure
seen in \eg\ the {\em Antennae} (a structure that appears more like a
non-collimated outflow originating from the galaxy-wide
star-formation, rather than a galactic wind-type outflow). One might
have expected the NGC520 wind structures to distort very quickly, due
to the rapid evolution and violence of the environment. However, the
structure appears, especially towards the centre, to be rather
classical in nature, and a single starburst-produced wind might indeed
last intact for some time during a major interaction.

\item 15 X-ray sources are detected within the optical confines of the
system. Of these, 3 or possibly 4 are ULXs, each with $L_{X}$
(0.2$-$10\,keV) \gtsim $10^{39}$\,erg s$^{-1}$. One of the ULXs lies
at the primary nucleus, at the centre of the diffuse wind structure,
and coincident with the bright radio source at the centre of the
molecular disk. A number of the other bright sources are seen to lie
near the edge of this molecular disk, in an interface region between
the progenitor-galaxy-driven centrally-located neutral ISM and the
merger-driven outlying ionised ISM. Significantly, no sources are
detected at the secondary nucleus, though within the system, much
apparent point-source variability is seen between the observations
described here and earlier {\em ROSAT} observations.

\item The slope of the source X-ray luminosity function (XLF) is very
flat, due predominantly to the ULXs, and appears in sharp contrast to
that of other normal and starburst galaxies. It is however, 
similar to that of the {\em Antennae} and other merging galaxies.

\item There is much in the X-ray data to support the idea, as
suggested by previous multi-wavelength studies, that NGC520 is the
result of an encounter between one gas-rich disk and one gas-poor
disk: The number of ULXs observed is only 3$-$4, while one would
expect, on the basis of the FIR luminosity, a number closer to
10. Similarly, $L_{\rm ULX}$ is lower than would be expected, based on
the $L_{\rm FIR}$ value. No sources are seen coincident with the
secondary (NW) nucleus, nor is any diffuse emission observed in or
around this nucleus. The total X-ray luminosity compared to other
multiwavelength luminosities, and the diffuse gas fraction are down by
about a factor of 2, based on what would be expected, given NGC520's
evolutionary merger stage. It does seem therefore, that, in terms of
the X-ray properties, because of the fact that NGC520's progenitors
are likely only one gas-rich disk plus one gas-poor disk, we are
observing only 'half a merger' when compared with similar
gas-rich$-$gas-rich merger systems, such as the {\em Antennae} and the
{\em Mice}.

\end{itemize}

\section*{Acknowledgements} 

AMR acknowledges the support of PPARC funding, and thanks the referee
(A.\,Zezas) for very useful comments which have helped to improve the
paper.


\end{document}